\documentclass[showpacs,oneside,superscriptaddress,pra,twocolumn]{revtex4}
\usepackage{amssymb}
\usepackage{amsfonts}
\usepackage{amssymb}
\usepackage{hyperref}
\usepackage{graphicx}

\begin{document}
\title{Experimental Realization of $1 \rightarrow 2$ Asymmetric Phase-Covariant Quantum Cloning}
\author{Hongwei Chen}
\affiliation{Hefei National Laboratory for Physical Sciences at
Microscale and Department of Modern Physics, University of Science
and Technology of China, Hefei, Anhui 230026, People's Republic of
China}
\author{Xianyi Zhou}
\affiliation{Hefei National Laboratory for Physical Sciences at
Microscale and Department of Modern Physics, University of Science
and Technology of China, Hefei, Anhui 230026, People's Republic of
China}
\author{Dieter Suter}
\affiliation{Fachbereich Physik, Universit\"{a}t Dortmund,
44221 Dortmund, Germany}
\author{Jiangfeng Du}
\email{djf@ustc.edu.cn} \affiliation{Hefei National Laboratory for
Physical Sciences at Microscale and Department of Modern Physics,
University of Science and Technology of China, Hefei, Anhui
230026, People's Republic of China} \affiliation{Fachbereich
Physik, Universit\"{a}t Dortmund, 44221 Dortmund, Germany}

\begin{abstract}
While exact cloning of an unknown quantum state is prohibited by the
linearity of quantum mechanics, approximate cloning is possible and has
been used, e.g., to derive limits on the security of quantum communication protocols.
In the case of asymmetric cloning, the information from the input state is
distributed asymmetrically between the different output states.
Here, we consider asymmetric phase-covariant cloning, where the goal is
to optimally transfer the phase information from a single input qubit to different
output qubits.
We construct an optimal quantum cloning machine for two qubits that does not require
ancilla qubits and implement it on an NMR quantum information processor.
\end{abstract}

 \maketitle

\section{Introduction}

Among the main differences between quantum and classical information
is the fundamental impossibility to exactly duplicate an unknown quantum state.
This was established by the no-cloning theorem of Dieks \cite{Dieks92}
and Wootters and Zurek \cite{Wootters299}; for a review, see Ref. \cite{Gisin242}.
While \emph{exact} cloning is thus impossible, it remains an important goal
to \emph{approximately} clone quantum states.
This possibility, which is particularly important for quantum communication and cryptography,
was first discussed by Bu\v{z}ek and Hillery \cite{Buzek54},
who showed that it is possible to create copies (approximate clones)
of unknown quantum states with a quality that does not depend on
the initial state.

Approximate cloning can be optimized in different ways.
In so-called asymmetric cloning, the amount of information transferred
from the input state to the copy is an adjustable parameter.
The quality of the copy and the distortion
that the cloning process causes on the original system both depend on
this parameter:
If the quality of the copy increases, the distortion of the original necessarily
increases simultaneously \cite{Cerf48,Cerf47,Cerf84,Niu58}.
This is quantified by the fidelity of the two output systems, which is defined
as the overlap of these states with the input state.
This tradeoff relates, e.g., the amount of information that an eavesdropper
can extract from a quantum communication channel to the error rate
of the transmitted information \cite{Cerf72}.

Asymmetric quantum cloning was first proposed for copying a single
qubit to a single copy qubit \cite{Cerf84}, and subsequently
extended to arbitrary dimensions (including the continuous case)
\cite{Braunstein63, Rastegin66}. An implementation of universal
asymmetric cloning in an optical experiment was proposed locally
\cite{Filip032309} and at a distance \cite{Filip052301}. An
experimental realization of  $1\rightarrow2$ asymmetric cloning
was reported by Zhao \emph{et al}. \cite{Zhao95} using two
entangled photon pairs.

The quality of the cloning can be improved if the initial state is restricted to part of the full
Hilbert space.
An example of this state-dependent cloning is phase-covariant cloning \cite{Bruss00},
where the input state is an equal-weight superposition of two basis states.
The goal is then to optimally clone the state in such a way that the phase information
is conserved.

In this paper, we construct a two-qubit quantum logic circuit that
implements the optimal asymmetric $1\rightarrow2$  phase-covariant
cloning \cite{Bruss00} for arbitrary input phase. Our cloning
machine does not require any ancilla qubits and uses only two gate
operations. The cloning process is implemented experimentally in
an NMR system, using nuclear-spin qubits. For the gate operations
we use controlled geometrical phase gates and demonstrate the
trade-off in fidelity for the two output qubits.

\section{Cloning scheme}

In the following, we consider phase-covariant cloning:
the original qubit to be cloned is in an equal-weight
superposition of the two basis states,
\begin{equation}
|\psi\rangle_{ini}=\frac{1}{\sqrt{2}}(|0\rangle+ \, e^{i\varphi}|1\rangle) ,
\end{equation}
with an unknown phase difference $\varphi$.
We clone this state onto a second qubit that is originally in state $|0\rangle$.
The cloning is acomplished by a unitary operation acting on the initial product state
$|\psi\rangle_{ini}|0\rangle$ \cite{Zhang356}.
The operation can be specified by its effect on the two orthogonal initial states
 $|00\rangle$ and  $|10\rangle$:
\begin{eqnarray}
&&|00\rangle\rightarrow|00\rangle,\nonumber\\
&&|10\rangle\rightarrow\cos\frac{\alpha}{2}|10\rangle+\sin\frac{\alpha}{2}|01\rangle
\end{eqnarray}
or by its matrix representation
\begin{equation}
\label{unitary}
u_{n}=\left(%
\begin{array}{cccc}
  1 & 0 & 0 & 0 \\
  0 & 0 & \sin\left(\frac{\alpha}{2}\right) & \cos\left(\frac{\alpha}{2}\right) \\
  0 & 0 & \cos\left(\frac{\alpha}{2}\right) & -\sin\left(\frac{\alpha}{2}\right) \\
  0 & -1 & 0 & 0 \\
\end{array}%
\right).
\label{e:un}
\end{equation}
This operation is equivalent (up to local operation and phases) to
$CROT_{12}CNOT_{21}$, and the rotation angle $\alpha$ can be used
to adjust how much information is transferred to the second qubit.

\begin{figure}[ht]
\includegraphics[width=7 cm]{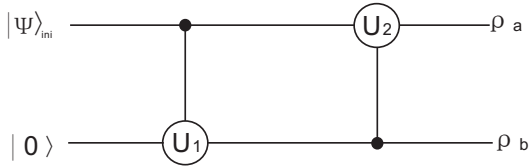}
\caption{The quantum logic circuit for our optimal asymmetric
quantum phase-covariant cloning machine. Qubit a is the one to be
cloned, initially in the state $|\psi\rangle_{ini}$, while qubit b
is the blank one which initially in state $|0\rangle$. The unitary
operation $U_{1}$ denotes $R_y(\alpha)$ and $U_{2}$ denotes
$R_y(-\pi)$.}
\end{figure}

After the cloning operation, the partial density operators for the two qubits are
$$
\rho_{a}= \frac{1}{2}\left(
\begin{array}{cccc}
 1+\sin^2\frac{\alpha}{2} &  e^{-i \varphi } \cos \frac{\alpha
   }{2}  \\
 e^{i \varphi } \cos \frac{\alpha }{2}  & 1-\sin^2\frac{\alpha}{2}
\end{array}
\right),
$$
\begin{equation}
\rho_{b}= \frac{1}{2}\left(
\begin{array}{cccc}
 1+\cos^2\frac{\alpha}{2} &  e^{-i \varphi } \sin \frac{\alpha
   }{2}  \\
 e^{i \varphi } \sin \frac{\alpha }{2}  & 1-\cos^2\frac{\alpha}{2}
\end{array}
\right) .
\end{equation}

The state-overlap between the original and the two output qubits are
\begin{eqnarray}
\label{fa}&&F_{a}=\rm{Tr}(\rho_{a}|\psi\rangle_{ini} \langle\psi|_{ini})
=\frac{1+\cos\frac{\alpha}{2}}{2},\nonumber \\
\label{fb}&&F_{b}=\rm{Tr}(\rho_{b}|\psi\rangle_{ini}\langle\psi|_{ini})
=\frac{1+\sin\frac{\alpha}{2}}{2}.
\end{eqnarray}
The choice of the angle $\alpha$ thus determines how much information is transferred
from qubit a to qubit b: For $\alpha=0$, the overlap of qubit a with the initial state
is 1, while the overlap of the copy qubit is just the random value $\frac{1}{2}$.
When $\alpha=\frac{\pi}{2}$, we obtain the case of optical symmetric cloning,
with $F_{a}=F_{b}\simeq0.85355$, which has been shown to be the optimal value
for symmetric phase-covariant cloning \cite{Bruss00}.
For $\alpha=\pi$, the information is transferred completely to the copy qubit,
with $F_{a}=\frac{1}{2},\,F_{b}=1$.

Compared to the logic circuit of the
symmetric cloning machine proposed in Ref \cite{Du94},  this
scheme needs fewer logic gates.
We therefore expect it to perform better in practice, being less affected
by experimental imperfections, such as errors in rotation angles of
radio-frequency pulses.

\section{Geometric phase gate}

Geometric quantum phases  \cite{Berry392,Aharonov58,Zhu61,Zhu85}
have the remarkable property that they depend only on global parameters
(e.g. the area of a circuit) and are therefore not sensitive to some
local variations of the trajectory.
It was therefore suggested, that quantum gate operations
using geometric gates may be less susceptible to experimental imperfections
and therefore yield higher fidelity
\cite{Zhu72,Zanardi264,Falci407,Duan292,Jones403,Wang87,Zhu89,Zhu66,Zhu67}

We therefore used geometric phase gates to implement the two controlled gate
operations required for the cloning operation (\ref{e:un}) (see Fig. 1).
We first discuss the relevant operation for a single qubit
and then extend the procedure to the controlled operations.

\subsection{Single qubit gate}

Within the two-level system, we consider two orthogonal states
$|\psi_{+}\rangle$ and $|\psi_{-}\rangle$, which undergo a cyclic evolution
described by the operator $U(\tau)$:
 $U(\tau)|\psi_{\pm}\rangle=e^{\pm i\gamma}|\psi_{\pm}\rangle$.
The parameter $\gamma$ is thus the total phase difference of the two states
acquired during this circuit.

In the computational basis ($|0\rangle$, $|1\rangle$), the cyclic states can be written as
$$
|\psi_{+}\rangle=\cos\frac{\chi}{2}|0\rangle+e^{i\varphi}\sin\frac{\chi}{2}|1\rangle ,
$$
$$
|\psi_{-}\rangle=\sin\frac{\chi}{2}|0\rangle-e^{i\varphi}\cos\frac{\chi}{2}|1\rangle ,
$$
where $(\chi,\varphi)$ are
the spherical coordinates of the state vector on the Bloch sphere.
For an arbitrary input state
$|\psi_{in}\rangle=a_{+}|\psi_{+}\rangle+a_{-}|\psi_{-}\rangle$
with $a_{\pm}=\langle\psi_{\pm}|\psi_{in}\rangle$, after the
cyclic evolution for the $|\psi_{\pm}\rangle$ state, the output
state is
$|\psi_{out}\rangle=U(\gamma,\chi,\varphi)|\psi_{in}\rangle$,where

\begin{equation}
U=
\left(%
\begin{array}{cc}
  e^{i\gamma}\cos^{2}\frac{\chi}{2}+ e^{-i\gamma}\sin^{2}\frac{\chi}{2} & ie^{-i\varphi}\sin\gamma\sin\chi \\
  ie^{i\varphi}\sin\gamma\sin\chi & e^{i\gamma}\sin^{2}\frac{\chi}{2}+ e^{-i\gamma}\cos^{2}\frac{\chi}{2} \\
\end{array}%
\right).
\end{equation}
This operation becomes a purely geometric gate operation if the dynamic
contribution to the total phase $\gamma$ vanishes.

\begin{figure}[ht]
\includegraphics[width=6 cm]{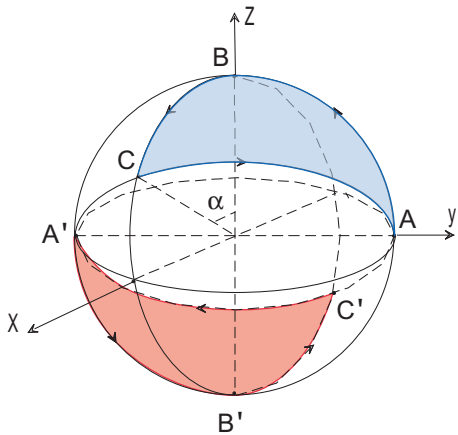}
\caption{\label{cy}The cyclic trajectories used in the experiment.
$|\psi_{+}\rangle$ is transported along the path A-B-C-A, and
$|\psi_{-}\rangle$ transported along the path A'-B'-C'-A'.}
\end{figure}

In the experiment, we use a nonadiabatic geometric phase and transform
the input states along geodesic circuits on the Bloch sphere,
as shown in Fig. 2.
Here, the cyclic states are
$$
|\psi_{+}\rangle = \frac{1}{\sqrt{2}}( |0\rangle+ i |1\rangle) ,\quad
|\psi_{-}\rangle = \frac{1}{\sqrt{2}}(|0\rangle- i |1\rangle ),
$$
i.e. $\chi=\varphi = \frac{\pi}{2}$.
For the circuit shown in Fig. 2, the solid angle subtended by the circuit is
equal to $\alpha$, the rotation angle during the second part of the circuit.
The geometric phase becomes thus $\gamma = -\frac{\alpha}{2}$.

For the circuit of Fig. 2, we may substitute the above values for
$\chi, \varphi,$ and $\gamma$.
The propagator becomes then
\begin{equation}
\label{uc}
U(-\frac{\alpha}{2},\frac{\pi}{2},\frac{\pi}{2})=
\left(%
\begin{array}{cc}
  \cos \frac {\alpha}{2} & \sin \frac{\alpha}{2}  \\
 - \sin \frac{\alpha}{2} & \cos \frac {\alpha}{2} \\
\end{array}%
\right).
\end{equation}

\subsection{Controlled operations}

We now apply this geometrical gate to controlled operations in a
two-qubit system where qubit a is the control qubit, while qubit b
undergoes the geometric circuit if qubit a is in state $|1\rangle$
but remains invariant if the control qubit is in state $|0\rangle$.

The Hamiltonian of the 2-qubit system is (in angular frequency units)
\begin{equation}
H = \omega_aI^a_z + \omega_b I^b_z + 2\pi J I^a_zI^b_z .
\label{e:Ham}
\end{equation}
For the subsystem of qubit b, we can write the reduced Hamiltonian
$$
H_b =   \omega_b I^b_z + 2 \pi J m_z^a I^b_z
 =  [ \omega_b - 2 \pi J (d^a -\frac{1}{2})] I^b_z   ,
$$
where $m_z^a$ is the eigenvalue of $I_z^a$ ($=\pm\frac{1}{2}$)
and $d^a$ the corresponding computational value ($=0,\,1$).

If we use a rotating frame with a frequency of $\omega_{b}'=\omega_{b} + \pi J$,
the Hamiltonian vanishes for $d^a=0$, $H_b^{(0)}=0$, while it becomes
$H_{b}^{(1)}=-2\pi J I^{b}_{z}$ for $d^a=1$.

This Hamiltonian generates controlled rotations around the z-axis.
To generate the trajectories of Fig. 2, we rotate the rotation axis
using radio-frequency pulses.
To generate a $\frac{\pi}{2}$ rotation around the $x$-axis, e.g.,
we use the sequence
$$
R_{y}^{b}(\frac{\pi}{2})-\frac{1}{4J}-R_{y}^{b}(-\frac{\pi}{2}) ,
$$
where the notation for the rotations is $R^{qubit}_{axis}(angle)$,
and $\frac{1}{4J}$ stand for free evolution under the control Hamiltonian
for the duration $\tau = \frac{1}{4J}$.
The circuit of Fig. 2 is thus generated by the sequence
\begin{eqnarray}
&&R_{y}^{b}(\frac{\pi}{2})-\frac{1}{4J}-R_{y}^{b}(-\frac{\pi}{2})-R_{x}^{b}
(-\frac{\pi}{2})-(\frac{\alpha}{2\pi J})\nonumber\\
&&-R_{x}^{b}(\frac{\pi}{2}) -R_{y}^{b}
(-\alpha-\frac{\pi}{2})-\frac{1}{4J}-R_{y}^{b}(\alpha+\frac{\pi}{2}).
\label{e:U1}
\end{eqnarray}

This represents the first gate operation of Fig. 1.
For the second operation, we have to reverse the roles of
control and target qubit and apply the following sequence to qubit a:
\begin{eqnarray}
&&R_{y}^{a}(\frac{\pi}{2})-\frac{1}{4J}-R_{y}^{a}(-\frac{\pi}{2})-R_{x}^{a}
(\frac{\pi}{2})-\frac{1}{2J}\nonumber\\
&&-R_{x}^{a}(-\frac{\pi}{2})-R_{y}^{a}(\frac{\pi}{2})
-\frac{1}{4J}-R_{y}^{a}(-\frac{\pi}{2}) ,
\label{e:U2}
\end{eqnarray}
now setting the rf frequency to $\omega_a + \pi J$.

\section{NMR Implementation}

\label{s:NMR}
For the experimental implementation, we used the two nuclear spins of
$^{13}$C-labelled chloroform as qubits.
The system Hamiltonian corresponds to Eq. (\ref{e:Ham}) with a spin-spin
coupling constant $J=214.5$ Hz.
Experiments were performed at room temperature on a Bruker AV-400
spectrometer.

The system was first prepared in a pseudopure state
$|00\rangle$ using the method of spatial averaging \cite{Cory120}
with the pulse sequence
\begin{equation}\label{inintial}
R^b_x (\pi/3 ) - G_z - R^b_x (\pi/4 ) - \frac{1}{2J} - R^b_y( \pi/4 ) -
G_z,
\end{equation}
which is read from left to right (as the following sequences).
The rotations  $R^{spins}_{axis}(angle)$ are implemented by radio-frequency pulses.
$G_z$ is a pulsed field gradient which
destroys all coherences (x and y magnetizations) and retains
longitudinal magnetization (z magnetization component) only.
$ \frac{1}{2J}$ represents a free precession period of the specified duration
under the coupling Hamiltonian (no resonance offsets).

From the state $|00\rangle$, we prepared the initial state
$\frac{1}{\sqrt{2}}(|0\rangle+e^{i\varphi}|1\rangle)\otimes|0\rangle$
by rotating qubit a (the $^{13}$C nuclear spin) into the $xy$-plane.
Experiments were done for  $\varphi=\frac{n\pi}{2}(n=0,1,2,3)$.
For each value of $\varphi$, we performed the cloning operation,
using the geometric gate operations (\ref{e:U1}) and  (\ref{e:U2})
for different asymmetry parameters $\alpha$.

To experimentally determine the fidelities (6), we need the density operators
of the initial state of qubit a and the final states of both qubits.
For this purpose, we parametrize the density operators as
$$
\rho = \frac{1}{2}(1 + \vec{r}\cdot \vec{I}) \, ,
$$
where $\vec{r}=(x,y,z)$ is a Bloch vector.
The fidelities (6) are then
$$
F_{i}=Tr(\rho_{0}\cdot\rho_{i})=\frac{1}{2}(1+\vec{r}_{0}\cdot \vec{r}_{i})),\,(i=a,b) ,
$$

For the initial state $\psi =
\frac{1}{\sqrt{2}}(|0\rangle+e^{i\varphi}|1\rangle)$, we have
$\vec{r}_0 = (\cos\varphi, \sin\varphi, 0)$ and the fidelities
become
\begin{equation}
F_i = \frac{1}{2} (1+ \cos\varphi x_i + \sin\varphi y_i) .
\label{e.fid}
\end{equation}
The transverse components $x_i, y_i$ can be measured as the
transverse magnetisation components of the
free induction decay.
To perform the trace over the other spin, we can either apply a decoupling field
to the other spin or integrate over the two lines in the spectrum.
For the present experiment, we chose the second possibility.

\section{Experimental results}

\begin{figure}[ht]
\includegraphics[width=9cm]{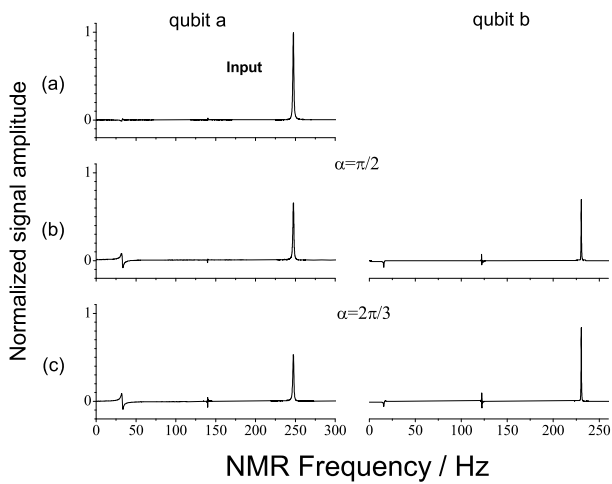}
\caption{\label{f.spectra}
Experimentally observed NMR spectra of $^{13}$C-chloroform before and
after a quantum cloning operation. The original state was an equal-weight
superposition with equal phases ($\varphi = 0$), shown in the top row.
The middle row shows the resulting spectra for a symmetric cloning operation
($\alpha = \pi/2$), and the bottom row the result of an asymmetric cloning
operation ($\alpha = 2\pi/3$).
The left hand column holds the input qubit, the right hand column
the copy qubit.
}
\end{figure}

Figure \ref{f.spectra} shows a typical example of a cloning operation.
The input qubit (qubit a) was initialized into a pseudo-pure state
$\frac{1}{\sqrt{2}}(|0\rangle+|1\rangle)$,
as described in section \ref{s:NMR}, using the phase angle $\varphi = 0$,
and the target qubit was set to $|0\rangle$.
This state corresponds to transverse magnetisation of spin $a$ and is therefore
directly observable in the NMR spectrometer.
The upper row of Figure  \ref{f.spectra} shows the Fourier transform
of the measured free induction decay (FID) of the $^{13}$C signal.
Only one of the two resonance lines is observable, indicating that the
target qubit is in the state $|0\rangle$.

The middle row of Figure \ref{f.spectra} shows the corresponding spectra
after a symmetrical cloning operation, with the propagator of Eq. (\ref{uc})
$$
U(\alpha=\frac{\pi}{2})=\frac{1}{\sqrt{2}} \left(
\begin{array}{cc}
  1 & -1  \\
1 & 1 \\
\end{array}%
\right).
$$
Integrating the signal for each spin species, we find for the x-components
$x_a = 0.667$ and $x_b = 0.682$, in good agreement with the
theoretical values of $x_a = x_b = 1/\sqrt{2}$.
The corresponding fidelities are
$F_a = 0.834$ and  $F_b = 0.841$ (theoretical values : 0.854).

The bottom row shows the same result for an asymmetric cloning operation.
Here, the rotation angle $\alpha$ was set to $2\pi/3$.
As a result, the target qubit has the higher fidelity:
$F_a = 0.758, F_b = 0.920$, again in good agreement with the
theoretical values of 0.750 and 0.933.

\begin{figure}[ht]
\includegraphics[width=8cm]{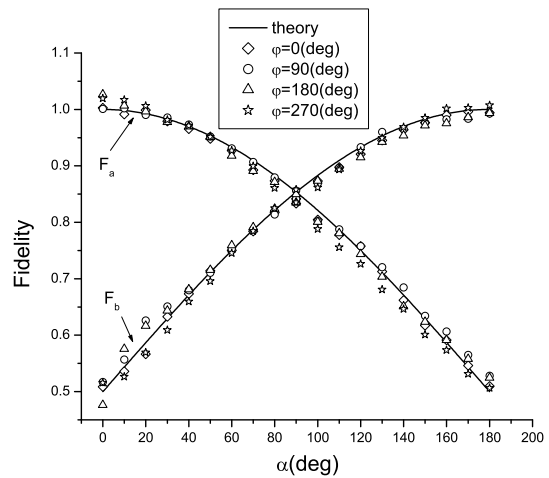}
\caption{\label{nf1}The experimental fidelities versus the
different parameter $\alpha$ of the asymmetric cloning machine.
The theoretical values of fidelities are plotted as solid lines.
And the different symbols are corresponding to the experimental
fidelities of two qubits with different angles $\varphi$ of the
initial state $|\psi\rangle_{ini}$.}
\end{figure}

Figure \ref{nf1} shows a more systematic check of the effect of the rotation angle
on the two fidelities.
We compare the fidelities of both qubits with the theoretical value
as a function of the asymmetry parameter $\alpha$.
The theoretical curve is independent of the phase
of the initial state.
Experimental data were measured for 4 different initial phases $\varphi$
as a function of the rotation angle $\alpha$.
All four data sets are in good agreement with the expectation.
For vanishing rotation, the input qubit is not disturbed
($F_a \approx 1$), while the target qubit bears no information
($F_b \approx 0.5$).
For a $\pi$-rotation, the roles of original and target qubit are reversed,
and at $\alpha = \pi/2$, both qubits share the information equally.

\begin{figure}[ht]
\includegraphics[width=8cm]{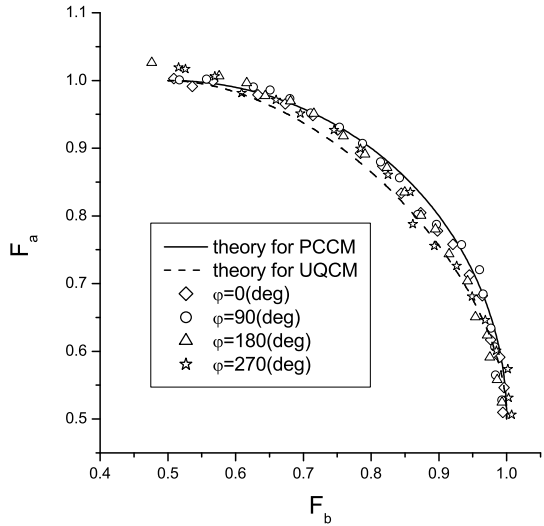}
\caption{\label{nf2}Trade-off diagrams in the asymmetric cloning
machine respectively for different phase angles $\varphi$ of the
initial state $|\psi\rangle_{ini}$. The full line shows the
theoretical expectation for phase-covariant cloning, while the
dashed line represents the limiting value for a universal cloning
machine. The different symbols refer to experimental data points
for different initial conditions. }
\end{figure}

This apparent complementarity of the two fidelities can be quantified.
According to Eqs.(\ref{fb}), the points $(F_a, F_b)$ are located on
a quarter-circle whose origin is at $(0.5, 0.5)$ and whose radius is 0.5.
Figure \ref{nf2} verifies this relation.
Here, the experimental fidelities are plotted against each other
for different rotation angles $\alpha$ and different initial phases
$\varphi$, represented by the different symbols.
All experimental points are close to the circle representing
the theoretical expectation \cite{Gia03}.
The dashed curve in Figure \ref{nf2} represents the theoretical
prediction for a universal cloning machine \cite{Bruss00}.
Evidently, the theoretical curve for phase-covariant cloning as well
as the experimental data are outside of this range,
except for angles close to 0 or $\pi$, where the information is located
on a single qubit.

\section{conclusion}

In summary, we have experimentally realized an optimal
asymmetric  $1\rightarrow2$ phase-covariant cloning machine.
As a function of a continuous angle variable in the cloning operation,
the phase information of the input state is transferred to the two output states
such that either the original qubit is only slightly disturbed ($\alpha \rightarrow 0$)
or that most of the phase information is transferred to the second qubit
 ($\alpha \rightarrow \pi$).
The case of symmetric cloning is recovered for $\alpha = \frac{\pi}{2}$.

In the case of quantum cryptography, this tradeoff determines how much
information the eavesdropper can gain for a given disturbance of the transmitted
information.
The fidelities found for this phase-covariant cloning were higher than
the upper bound for universal cloning.

The cloning was implemented experimentally on an NMR quantum information processor.
For the cloning operations, we used cyclic rotations of the qubits
in such a way that the system acquired a geometrical phase.
This procedure has been proposed for shielding the gate operation from
such perturbations that leave the area of the quantum mechanical trajectory invariant
and thereby improve the overall fidelity.

\acknowledgments
We thank Prof. Z. D. Wang and Dr. Q. Chen for helpful discussions.
This project is supported by the National Fundamental Research
Program (Grant No. 2001 CB 309300),
NSFC under Grant No. 10425524 and No. 10429401,
the National Science Fund of China,
and the European Commission under Contract No. 007065 (Marie Curie).


\begin{thebibliography}{99}
\bibitem{Dieks92}D. Dieks,  Phys. Lett. A  \textbf{92}, 271(1982).
\bibitem{Wootters299}W. K. Wootters and W. H. Zurek,  Nature(London) \textbf{299}, 802(1982).
\bibitem{Gisin242}V. Scarani, S. Iblisdir, N. Gisin and A. Acin,  Rev. Mod. Phys. \textbf{77}, 1225 (2005).
\bibitem{Buzek54}V. Bu\v{z}ek and M. Hillery,  Phys. Rev. A  \textbf{54}, 1844(1996).
\bibitem{Cerf48}N. J. Cerf, Acta Phys. Slov. \textbf{48}, 115(1998).
\bibitem{Cerf84}N. J. Cerf, Phys. Rev. Lett. \textbf{84}, 4497(2000).
\bibitem{Cerf47}N. J. Cerf, Mod. Opt. \textbf{47}, 187(2000).
\bibitem{Niu58}C. S. Niu and R. B. Griffiths, Phys. Rev. A  \textbf{58}, 4377(1998).
\bibitem{Cerf72}S. Iblisdir, A. $Acin$ and N. J. Cerf \emph{et al.}, Phys. Rev. A \textbf{72}, 042328(2005).
\bibitem{Braunstein63}S. L. Braunstein, V. Bu\v{z}ek and M. Hillery, Phys. Rev. A \textbf{63}, 052313(2001).
\bibitem{Rastegin66}A. E. Rastegin, Phys. Rev. A \textbf{66}, 042304(2002).
\bibitem{Filip032309}R. Filip, Phys. Rev. A \textbf{69}, 032309(2004).
\bibitem{Filip052301}R. Filip, Phys. Rev. A \textbf{69}, 052301(2004).
\bibitem{Zhao95}Z. Zhao, A. N. Zhang and X. Q. Zhou \emph{et al.}, Phys. Rev. Lett. \textbf{95}, 030502(2005).
\bibitem{Bruss00}D. Bruss, M. Cinchetti, G. M. d'Ariano and C. Macchiavello, Phys. Rev. A. \textbf{62}, 012302(2000).
\bibitem{Zhang356}W. H. Zhang, L. B. Yu and L. Ye,  Phys. Lett. A  \textbf{356}, 195(2006).
\bibitem{Du94}J. F. Du, T. Durt and P. Zou \emph{et al.}, Phys. Rev. Lett. \textbf{94}, 040505(2005).
\bibitem{Cory120}D. G. Cory, M. D. Price and T. F. Havel, Phys. D \textbf{120}, 82(1998).
\bibitem{Zhu72}S. L. Zhu and P. Zanardi, Phys. Rev. A \textbf{72}, 020301(2005).
\bibitem{Berry392}M. V. Berry, Proc. R. Soc. London, Ser A \textbf{392}, 45(1984).
\bibitem{Aharonov58}Y. Aharonov and J. Anandan, Phys. Rev. Lett. \textbf{58}, 1593(1987).
\bibitem{Zhu61}S. L. Zhu, Z. D. Wang and Y. D. Zhang, Phys. Rev. B  \textbf{61},1142(2000).
\bibitem{Zhu85}S. L. Zhu and Z. D. Wang, Phys. Rev. Lett. \textbf{85}, 1076(2000).
\bibitem{Zanardi264}P. Zanardi and M. Rasetti, Phys. Lett. A  \textbf{264}, 94(1999).
\bibitem{Falci407} G. Falci, R. Fazio and G. M. Palma \emph{et al.},  Nature(London) \textbf{407}, 355(2000).
\bibitem{Duan292} L. M. Duan, J. I. Cirac and P. Zoller,  Science \textbf{292},1695 (2001).
\bibitem{Jones403} J. A. Jones, V. Vedral, A. Ekert and G. Castagnoli,  Nature(London) \textbf{403}, 869(2000).
\bibitem{Wang87}X. B. Wang and M. Keiji, Phys. Rev. Lett. \textbf{87}, 097901(2001).
\bibitem{Zhu89}S. L. Zhu and Z. D. Wang, Phys. Rev. Lett. \textbf{89}, 097902(2002).
\bibitem{Zhu67}S. L. Zhu and Z. D. Wang, Phys. Rev. A \textbf{67}, 022319(2003).
\bibitem{Zhu66}S. L. Zhu and Z. D. Wang, Phys. Rev. A \textbf{66}, 042322(2002).
\bibitem{Zhang71}X. D. Zhang, S. L. Zhu, L. Hu and Z. D. Wang, Phys. Rev. A \textbf{71}, 014302(2005).
\bibitem{Du74}J. F. Du, P. Zou and Z. D. Wang, Phys. Rev. A \textbf{74}, 020302(2006).
\bibitem{Gia03}Giacomo Mauro D'Ariano and Chiara Macchiavello, Phys. Rev. A. \textbf{67}, 042306(2003).


\end{thebibliography}
\end{document}